\documentclass[12 pt]{article}
\usepackage{times}
\usepackage{subfigure}

\usepackage{pifont}
\usepackage{floatrow}
\usepackage{caption}

\usepackage{mathrsfs}

\usepackage[fleqn]{amsmath}
\usepackage{amsfonts,amsthm,amssymb,mathrsfs,bbding}
\usepackage{txfonts}
\usepackage{graphics,multicol}
\usepackage{graphicx}
\usepackage{color}
\usepackage{caption}
\usepackage{indentfirst}
\usepackage{cite}
\usepackage{latexsym,bm}
\usepackage{enumerate}
\usepackage{epsfig}
\usepackage[noend]{algpseudocode}
\usepackage[ruled,vlined]{algorithm2e}
\usepackage{tikz}
\usepackage{hyperref,todonotes,ulem,comment}
\usepackage{empheq,cases}

\newtheorem{remark}{Remark}

\def\x{\bm{x}}

\def\y{\bm{y}}
\def\z{\bm{z}}

\def\y{\bm{y}}
\def\btheta{\bm{\theta}}
\def\@#1{{\cal #1}}

\begin{document}
\setlength {\marginparwidth }{2cm}
\title{Adaptive design of experiments via normalizing flows for failure probability estimation}

\author{Hongji Wang$^a$,\quad Tiexin Guo$^a$, \quad Jinglai Li$^{b}$
{and} \quad
Hongqiao Wang$^{a*}$ \\
{\it $^a$School of Mathematics and Statistics} \\
{\it Central South University} \\
{\it Changsha 410083, People’s Republic of China}\\
[2mm]
{\it $^b$School of Mathematics} \\
{\it  University of Birmingham} \\
{\it  Edgbaston, Birmingham B15 2TT, UK}\\
[2mm]
{$^*$ Corresponding author: Hongqiao Wang}\\
{\it School of Mathematics and Statistics} \\
{\it Central South University} \\
{\it Changsha 410083, People’s Republic of China}\\
{\it E-mail: Hongqiao.Wang@csu.edu.cn}
 }

\date{}

\maketitle {\flushleft\large\bf Abstract }

Failure probability estimation problem is an crucial task in engineering.
In this work we consider this problem in the situation that the underlying computer models are extremely expensive, which often arises in the practice, and in this setting, reducing the calls of computer model is of essential importance.
We formulate the problem of estimating the failure probability with expensive computer models as an sequential experimental design for the limit state (i.e., the failure boundary) and propose a series of efficient adaptive design criteria to solve the design of experiment (DOE). 
In particular, the proposed method employs the deep neural network (DNN) as the surrogate of limit state function for efficiently reducing the calls of expensive computer experiment.
A map from the Gaussian distribution to the posterior approximation of the limit state is learned by the normalizing flows for the ease of experimental design.
Three normalizing-flows-based design criteria are proposed in this work for deciding the design locations based on the different assumption of generalization error.
The accuracy and performance of the proposed method is demonstrated by both theory and practical examples.



\section{Introduction}
Real-life engineering systems are unavoidably subject to various uncertainties
such as material properties, geometric parameters,
boundary conditions and applied load.
These uncertainties may cause undesired events, in particular, system failures or malfunctions, to occur.
Accurate identification of failure region and evaluation of
failure probability of a given system are essential tasks in many engineering fields such as risk
management, structural design,  {reliability-based} optimization, etc.

Conventionally the failure probability is often computed by constructing linear or quadratic expansions of the system model around
the so-called most probable point, known as the first/second-order reliability method~(FORM/SORM), see e.g., \cite{SCHUELLER2004463}
and the references therein.
It is well known that FORM/SORM may fail for systems with  {nonlinearity} or multiple failure regions.
The Monte Carlo (MC) simulation, which estimates the failure probability by repeatedly simulating the underlying system, is another popular method for solving such problems. The MC method makes no approximation to the underlying computer models and thus can be applied to any systems.
On the other hand, the MC method is notorious for its slow convergence,
and thus can become prohibitively expensive
 when the underlying computer model is
computationally intensive and/or the system failures are rare  {and each sample} requires a full-scale
numerical simulation of the system.
To reduce the computational effort,  one can construct an computationally inexpensive approximation of the true model, and then evaluate
the approximate model in the MC simulations.  Such approximate models are also known as response surfaces,
surrogates, metamodels,  {and emulators, etc. These methods} are referred to as
 the response surface (RS) methods~\cite{faravelli1989response,gayton2003cq2rs,oakley2002bayesian} in this work.
 The response surface  can often provide a reliable estimate of the failure probability, at a much lower computational cost than direct MC simulations.

In this work we are focused on a specific kind of RS, the deep neural network (DNN) surrogates.
The DNN surrogates have been widely used in machine learning~\cite{liu2017survey}, geostatistics
~\cite{wang2019nearest},  engineering optimizations~\cite{abd2021advanced}, and most recently, uncertainty quantifications~\cite{tripathy2018deep,kabir2018neural}. 
In this work we consider the situation where the underlying computer models are extremely expensive and one can only afford a very limited number of simulations.
In this setting, choosing the sampling points (i.e. the parameter values with which the simulation is performed) in the state space is of essential importance.
Determining the sampling points for neural network can be cast as to optimally design computer experiments.
A simple and straightforward idea aims to construct a surrogate that can accurately approximate the target function in the whole parameter space.
As will be explained later, in the failure probability estimation or failure detection problems, only the sign of the target function is used. 
Thus with requiring surrogates to be globally accurate, the method may allocate considerable computational efforts to the regions not of interest,
and use much more model simulations than necessary.

Several methods have been developed to determine the sampling points for the failure probability estimation.
Most of these methods consist of sequentially finding the "best point" as a result of a heuristic balance between predicted closeness to the limit state, and high prediction uncertainty, e.g. \cite{echard2011ak,bichon2008efficient}.
Such methods are shown to be effective in many applications, while
a major limitation is their point-wise  {sequential} nature, which  makes it  unsuitable for problems in which multiple computer simulations can be performed parallelly.
The stepwise uncertainty reduction (SUR) method developed in \cite{bect2012sequential,chevalier2014fast}  is one of the two exceptions, in which the authors proposed an optimal experimental design framework which determines multiple sampling points by minimizing the average variance of the failure probability.
It should be noted that the design criteria in the SUR method is particularly developed for the goal of estimating the failure probability only.
In practice, one is often not only interested in estimating the failure probability, but also identifying the events that can cause failures;
the latter demands a design criteria for the goal of detecting the limit state, i.e., the boundaries of the failure domain.
Another exception is the Gaussian process based failure boundary and probability estimation methods\cite{wang2016gaussian,renganathan2022multifidelity}, which determine the multiple sampling points by maximizing  the information gain based design criteria.
The above multiple sampling points design methods would suffer from the same bottleneck that the difficulty of optimization would increase as the amount and the dimension of design points increase.
This bottleneck seriously reduces the possibility of searching global optimal design and limits the application of these methods in real world.
In this work, we recast the neural network (NN) surrogate construction as a Bayesian inference to identify the distribution of limit state, and based on that, we propose three normalizing-flows-based design criteria to determine the sampling points.
The proposed neural network based method could be easily applied in the high dimensional case and it avoids the step of optimization which is essential in the above optimal design methods.
We compare the performance of the proposed method with that of the LSI by numerical examples.


We note that another line of research in failure probability estimation is  {to} develop more efficient sampling schemes, such as the subset simulations~\cite{au2001estimation}, importance sampling~\cite{engelund1993benchmark}, the cross-entropy method~\cite{rubinstein2004applications,wang2015cross}, etc.
For practical engineering systems, computer simulations can be
extremely time consuming. In many cases, one can only afford very
limited number of simulations. 
In this case, even the most effective sampling method is not applicable.
To this end, surrogates are needed even in those advanced sampling schemes and in particular the proposed method can be easily integrated into the aforementioned sampling schemes,
resulting in more efficient estimation schemes. Examples of combining surrogates and efficient sampling schemes include \cite{li2011efficient,li2012bayesian,dubourg2013metamodel}.

The rest of this paper is organized  {as following}. We first review
the preliminaries of our work in Section
\ref{set:problem}, including the mathematical formulation of failure probability computation and the DNN surrogates. Our sequential failure probability estimation framework and its numerical implementations are presented in Section~\ref{set:sequ_FP}. 
Numerical examples are presented in Section \ref{set:numerical_examples} to demonstrate the
effectiveness of the proposed method, and finally  {Section~\ref{set:conclusion}} offers some closing remarks.

\section{Problem formulation}
\label{set:problem}
\subsection{Failure probability estimation framework}
\label{set:failure_prob}

In a general setting of failure probability estimation problem~\cite{wang2016gaussian}, 
we consider  a $d$-dimensional random variable $\x$  that represents input variable with uncertainty and let $\Omega \subseteq \mathbb{R}^d$ be the state space of $\x$.
Let $(\Omega, \mathcal{F}, \mathbb{P})$ be a probability space, where $\Omega$ is a sample space, $\mathcal{F}$ is a $\sigma$-field, and $\mathbb{P}$ is a probability measure on $(\Omega, \mathcal{F})$.
We model a system using a real-valued function $g(\cdot):\Omega \to R$, which is known as the limit state function or the performance function. 
The event of failure is defined as $g(\x)\le 0$ and as a result the failure probability is 
\begin{equation}
    P = \mathbb{P}(g(\x)\le 0) = \int\limits_{\{\x\in \Omega|g(\x)\le 0\}} p_X(\x)d\x = \int\limits_{ \Omega}I_g(\x)p_X(\x)d\x,
\end{equation}
where $I_g(\x)$ is an indicator function:
\begin{equation}
    I_g(\x)=\Big\{ \begin{array}{cc}
        1 & \text{if } g(\x) \le 0,  \\
        0 & \text{if } g(\x)> 0;
    \end{array}
\end{equation}
and $p_X(\x)$ is the probability density function (PDF) of $\x$. 
In what follows we shall omit the integration domain when it is simply $\Omega$.
This is a general definition for failure probability, which is used widely in many disciplines involving reliability analysis and risk management.
$P$ can be computed with the standard Monte Carlo (MC) estimation:
\begin{equation}
\label{eq:MC}
    \hat{P} = \frac{1}{n}\sum^n_{i=1} I_g(\x_i),
\end{equation}
where samples $\x_1,\dots,\x_n$ are drawn from $p_X(\x)$ which can be any probability density function.
The failure probability can be estimated by MC method and this method does not require any assumptions on it.

A high reliable estimate of the small failure probability, for example, $P\ll 1$, is required in many practical engineering systems. 
In this case, MC requires a rather large number of samples to produce a reliable estimate of the failure probability. For example, for $P\approx 10^{-3}$, MC simulation requires $10^5$ samples to obtain an estimate with $10\%$ coefficient of variation.
On the other hand, in almost all practical cases, the limit state function
$g(\x)$ does not admit analytical expression and has to be evaluated through expensive computer simulations, which renders the crucial MC estimation of the failure probability prohibitive. 
To reduce the number of full-scale computer simulations, one can construct a computationally inexpensive surrogate $G(\x)$ to replace the real function $g(\x)$ in the MC estimation. In this work we choose the powerful Deep Neural Network (DNN) model as an efficient surrogate $G(\x)$ due to its outstanding success in high dimensional scenario.

\subsection{Deep Neural network surrogate of expensive computer simulation}
\label{set:DNN}
The powerful DNN  have shown its strong fitting ability in lots of areas\cite{jumper2021highly,raissi2019physics}.
Here we employ it to construct the surrogate of real expensive limit state function $g(\x)$.
With generality, we prefer to use the full connection Neural Network architecture which is the most simple Neural network structure as the surrogate function $G(\cdot):=G_{out}\circ \dots \circ G_l\circ \dots \circ G_{in}(\cdot)$ in most cases.
But for more challenging limit state functions like parametric PDE solvers, we prefer a  more complex architecture, like Fourier Neural Operator (FNO)~\cite{li2020fourier}.

The basic idea of deep neural networks (DNNs) for surrogate model is that it can approximate an input-output map $G :\mathbb{R}^d \to\mathbb{R}$ through a hierarchical abstract layers of latent variables.
A typical example is the feedforward neural network, which is also called multi-layer perception (MLP). It consists of a collection of layers that include an input layer $G_{in}(\cdot)$, an output layer $G_{out}(\cdot)$, and a number of hidden layers $G_k(\cdot), k=1,\dots,K$. 
The size of the input layer and output layer are fixed and determined by the dimensionality of the input and output.
Each element of $G_k(\cdot)$ is a neuron which calculates a
weighted sum of an input vector plus bias and applies a non-linear function to produce an output.
Specifically, given an $d$-dimensional input row vector $\x\in \mathbb{R}^d$, we can define a DNN with $L$ hidden layers as following
\begin{align}
    \mathcal{NN}(\x)&=W^{(L)}a^{(L)} + b^{(L)}\\
    a^{(k+1)} &= \sigma(W^{(k)}a^{(k)} + b^{(k)}),\quad k=0,\dots,L-1.
\end{align}
Here $W^{(k)}\in \mathbb{R}^{d_{k+1}\times d_k}$, $b^{(k)}\in \mathbb{R}^{d_{k+1}}$ are the weights and biases of the network,
$d_k$ is the number of neurons in the $k$th layer and $\sigma$ is the activation function.
Notice that here $a^{(0)}$ is the input $\x$ and $d_0=d$. 
Some popular choices for the activation function include sigmoid,
hyperbolic tangent, rectied linear unit (ReLU), to name a few.
In the current work, we shall use Swish as the activation function:
\begin{equation}
    \sigma(z)=\frac{z}{1+\exp(-z)}.
\end{equation}

Once the network architecture is defined, one can resort to optimization tools to find the unknown parameters $\btheta = \{W^{(k)}, b^{(k)}\}$ based on the training data.
Precisely, let $\mathcal{D}:=\{ (\x_i,y_i)\}^N_{i=1}$ be a set of training data, we can define the following minimization problem:
\begin{equation}
    \arg \min \mathcal{J}(\btheta;\mathcal{D})= \frac{1}{N}\sum^N_{i=1} ||y_i - \mathcal{NN}(\btheta;\x_i) ||^2,
\end{equation}
where $\mathcal{J}(\btheta;\mathcal{D})$ is the so called loss function.
Solving this problem is generally achieved by the stochastic gradient descent (SGD) algorithm
which minimizes the function by taking a negative step along an estimate of the gradient
$\nabla$ at iteration $k$. The gradients are usually computed through back propagation.
At each iteration, SGD updates the solution by
\begin{equation}
    \btheta_{k+1} = \btheta_k - \epsilon\nabla_\theta\mathcal{J}(\btheta;\mathcal{D}_M),
\end{equation}
where $\epsilon$ is the learning rate and $\mathcal{D}_M$ is batch dataset.
Recent algorithms that offer adaptive learning rates are
available, such as Ada-Grad \cite{lydia2019adagrad}, Adam and RMSProp \cite{zou2019sufficient}, ect. The present work adopts Adam optimization algorithm.

For parametric PDE problem, we could employ a more efficient neural network architecture, named as Fourier Network Operator (FNO) \cite{li2020fourier}, to construct an more complex and accurate surrogate. 
The last example in Section \ref{set:numerical_examples} will introduce the failure probability estimation problem in PDE situation in detail.

\section{The sequential failure probability estimation framework}
\label{set:sequ_FP}
The failure probability can be estimated by the NN surrogate under the assumption that the data points are determined all in advance of performing computer simulations, which is often referred to as an open-loop design. 
But an accurate NN surrogate needs to be trained with a large number of data which is still computational intensive.
In many applications, a more practical strategy is to choose the sampling points in a sequential fashion: determine a set of sampling points, perform simulations, determine another set of points based on the previous results, and so forth. 
A sequential (close-loop) design can be readily derived from the open-loop version. 
Simply speaking, the sequential design iterates as follows until a prescribed stopping criterion is met:
\begin{itemize}
	\item[1] construct a NN model $G(\x)$ for $g(\x)$ using data-set $\mathcal{D}$;
    \item[2] determine $n$ sampling points $\{ \x^*_1,\dots,\x^*_n \}$ with an open-loop design;
    \item[3] evaluate $y^*_i = g(\x^*_i), i=1,\dots,n$ and let $\mathcal{D} = \{ \mathcal{D},(\x^*_1,y^*_1),\dots,(\x^*_n,y^*_n)\}$;
\end{itemize}
Note that the key in the sequential scheme is step $2$, where we efficiently seek the more informative sampling points.
Different from the traditional optimization-based experimental design criteria \cite{bect2012sequential,chevalier2014fast,wang2016gaussian,renganathan2022multifidelity},
here we propose a series of novel normalizing-flows-based sampling strategies which map an common distribution to the posterior distribution of limit state and decide the design points based on the samples of it in Section \ref{set:NFBD}. 
These criteria avoid the challenge of searching the global optimal which is a major problem in the field of optimization and reduces the undetermined time required for optimization.
Before introducing the specific design criteria, we first define  the posterior distribution of limit state which is an important concept in our method for the estimation of failure probability.

\subsection{Posterior distribution of limit state }
In the failure probability estimation, the limit state function $g$ is only used in the indicator function $I_g(\x)$ and so one is really interested in the sign of $g(\x)$ rather than the precise value of it. 
To this end, the essential task in constructing surrogate for the failure probability estimation is to learn about the boundary of the failure domain. Here we emphasize that the indicator function $I_g(\x)$ is a step function, but the limit state function $g(\x)$ is continuous.
Let $Z = \{\z:=\x|\x\in \Omega, g(\x)=0 \}$
represents the boundary of the failure domain, i.e., the collection of solutions of $g(\x)=0$
and define the distribution of $\z$, $p(\z) = \frac{1}{C}\exp(-\frac{|g(\z)-0|}{\lambda})$, where $C$ is a normalization constant.
Similarly, we define the posterior distribution of limit state in surrogate, i.e.,
\begin{equation}
\label{eq:post_limit_state}
    p(\tilde{\z}| \mathcal{D}):=p(G(\x)=0| \mathcal{D}) \propto \exp(-\frac{|G(\x)-0|}{\lambda}),
\end{equation} 
where $\lambda$ is scale parameter which scales the magnitude the output $G(\x)$ and $G(\x)$ follows the description of Section \ref{set:DNN}.
With the increment of data, the posterior distribution of limit state in surrogate $p(\tilde{\z}|\mathcal{D})$, would converge to the $p(\z)$.

\subsection{Density transformation via normalizing flows}
\label{set:NF}
Though we can obtain the unnormalized posterior density of limit state, we prefer calling the true posterior density and its samples. 
Here we employ the normalizing flows technique to approximate the limit state posterior which have been shown its strong fitting ability in \cite{rezende2015variational}.
NF attracts us with its cheap sampling procedure and density calculation.

The basic rule for transformation of densities considers an invertible, smooth map $f:\mathbb{R}^d\to \mathbb{R}^d$ with inverse $f^{-1}=h$, i.e. the composition $h \circ f(\z_0) = \z_0$. 
If we use this map to transform a random variable $\z_0$ with distribution $q(z_0)$, the resulting random variable $\z=f(\z_0)$ has a distribution:
\begin{equation}
\label{eq:NF_flow}
q(\tilde{\z})=q(\z_0)|\det \frac{\partial f^{-1}}{\partial \tilde{\z}}| = q(\z_0)|\det \frac{\partial f}{\partial \z_0}|^{-1},
\end{equation}
where the last equality can be seen by applying the chain rule (inverse function theorem) and the property of Jacobians of invertible functions.
DNN is used to exactly approximate $f$ for constructing arbitrarily complex densities.
For completeness of the paper, we briefly introduce the normalizing flows in Appendix \ref{app:NF}.

\subsection{Adaptive experimental design via normalizing flows}
\label{set:NFBD}
\textbf{normalizing-flows-based design (NFBD)} 

A simple and common adaptive strategy for experimental design is to take the experiments at the locations where we are interested \cite{papamakarios2019sequential,chen2019adaptive,blum2010non}.
In this problem we are interested in the distribution of limit state, $p(\tilde{\z}|\mathcal{D})$, and therefore a basic adaptive experimental design criterion is to use the sample points of $p(\tilde{\z}|\mathcal{D})$ as the design locations.
We could obtain the transform map $f$ by normalizing flows in section \ref{set:NF} and then the sample points (design locations) of $p(\tilde{z}|\mathcal{D})$, can be drawn easily by the transformation $\tilde{\z}=f(\z_0)$, where $\z_0$ are the samples of Gaussian distribution $p(\z_0)$.

\textbf{normalizing-flows-based design with fixed generalization (NFBD-FG)}

The basic adaptive sampling strategy is effective, but would generate some less informative designs since there will be a lot of experiments lying in the high probability region of limit state and some of them would flock together.
Two adjacent designs will provide duplicate information which is meaningless for improving the accuracy of surrogate.
With the consideration of the data sparsity, interpolation uncertainty here is an inherent epistemic uncertainty associated with machine learning models, when they are used to predict new data points through interpolation/extrapolation.
Here we give a fixed smoothness assumption about the limit state function and a straightforward strategy of reducing experimental calls can be obtained by adding a selection procedure  on the NFBD criterion, named as NFBD-FG.
For measuring the distance between a design candidate $\z$ and points in the current data-set $\mathcal{D}$, we define a distance function 
\begin{equation}
\label{eq:rho_x_D}
\rho(\z,\mathbb{D})=\min(|| \z-\x_1||_{L_{\infty}},\dots,|| \z-\x_m||_{L_{\infty}}),
\end{equation}
where $\x_i$ represents the experiment location in $\mathcal{D} = \{(\x_i,y_i)\}, i=1,\dots,m$.
This distance function could return the minimum distance value between the design candidate and data locations of current data-set $\mathcal{D}$.
Here we sequentially measure the distance between each proposed design candidate and the exiting experimental locations and accept the proposed sample with $1_{\rho(x,\mathbb{D})\le \epsilon}$, where $\epsilon$ is a preset value indicating a fixed generalization assumption.

\textbf{normalizing-flows-based design with adaptive generalization error (NFBD-AG)}

Though we can simply use the distance function $\rho(\cdot,\cdot)$ to quantify the interpolation error of surrogate, the impact of it would be different in the input variable domain, i.e., the failure probability error caused by same interpolation error in high probability domain would be smaller than the one in low probability domain.
Thus we propose an adaptive threshold strategy: the threshold value would not be constant but a function with respect to the probability density function value of input variable at proposed design points.
It has been shown that when the expectation is calculated over all possible data distributions, as sample complexity increases, generalization error will decline following a power-law~\cite{wang2022generalization}
\begin{equation}
\label{eq:adaptive_epsilon}
e(N)\sim \alpha N^{\gamma}.
\end{equation} 
This pow-law rule was observed in some empirical studies. 
It is worth noting that there are many choices of $\gamma$ in real applications and in this paper we choose $\gamma = -0.5$ without loss of generality.
In our generalization set as shown in Eq. \eqref{eq:rho_x_D} and the rule of acceptance $1_{\rho(x,\mathbb{D})\le \epsilon}$, we have $N = \frac{1}{\epsilon^d}$.
Eq.\eqref{eq:adaptive_epsilon} becomes 
\begin{equation}
\label{eq:adaptive_epsilon2}
e(\epsilon)\sim \alpha (\frac{1}{\epsilon^d})^{-0.5}.
\end{equation} 
We note that more samples draw from the high probability domain of $p(\x)$ for the computation of failure probability in MC methods and thus the same generalization error in different probability area of $p(\x)$ would lead to different error bounds in the estimation of failure probability. 
For alleviating the above difficulty, we propose an adaptive generalization strategy in Equation \eqref{eq:AG} which means the generalization error is not set up as a fixed value in advance, but adaptively adjusted with the probability of $p(\x)$, named as NFBD-AG.
The adaptive generalization error method could decline the generalization error in high probability area and allow a larger generalization error in low probability area.
\begin{align}
\label{eq:AG}
e(\epsilon(\x))&\propto \frac{1}{p(\x)} \Rightarrow \epsilon(\x) \propto \beta \frac{1}{p(\x)^{\frac{2}{d}}}
\end{align}
where $\beta = \frac{1}{\alpha^{\frac{d}{2}}}$.

\begin{remark}
In practice we could not know the exact value of $\beta$.
Here we set a empirical preset threshold $\epsilon_0$ corresponding to the median pdf value of $p_X(\z)$ in each iteration and then we have $\beta = \epsilon_0p(\z_{median})^{\frac{2}{d}}$.
For preventing the value $\epsilon(\z)$ too extreme, a bound $[0.1\epsilon_0,10\epsilon_0]$ is used here to constrain the adaptive threshold $\epsilon(\z)$ in a reasonable range, not too big (larger than $10\epsilon_0$) or too small (smaller than $0.1\epsilon_0$).

\end{remark}

\subsection{Numerical implementation}
\label{set:alg}
\textbf{Design via normalizing flows (DNF) algorithm}\\
For clearly illustrating our proposed design via normalizing flows method, named as "DNF", and list the full algorithm as follows:\\
 \IncMargin{1em}
\begin{algorithm}[H]\label{alg}
\caption{DNF method}
\LinesNumbered 
\SetKwInOut{Input}{Parameter}\SetKwInOut{Output}{Output}
\SetAlgoLined
\Input{max number of simulations $N_{\max}$, number of initial data points $N_0$, number of design locations  
in each adaptive experimental design update $N_D$ and the default threshold $\epsilon_0$;} 
\Output{estimation of failure probability $P_f$; }
Randomly select $N_0$ points $(\x_1,\dots,\x_{N_0})$ and compute $y_i=g(\x_i)$. Obtain the initial data-set  $\mathcal{D} = \{ \x_i, y_i\}_{i=1,\cdots,N_0}$\;
Train the NN surrogate  function $G$ and get the model parameters $\theta$ by minimizing the loss function with the data-set $\mathcal{D}$  (Section \ref{set:DNN})\;
Estimate the failure probability $\hat{p}_f^0$ by Equation \eqref{eq:MC} with the NN surrogate $G(\x)$ and compute $t_{\max} = \lceil\frac{N_{max}-N_0}{N_D}\rceil$\;
\For{$t=1$ \KwTo$t_{\max}$} 
{
\eIf {$t<t_{max}$}{$N_t = N_D$;}{$N_t = N_{\max} - N_0 - (t_{max}-1)*N_D$}
Train the normalizing flows and obtain the map from a common distribution $p(\z_0)$ to the posterior distribution of limit state $p(\tilde{z}|\mathcal{D})$ defined in \eqref{eq:post_limit_state}\;
Determine $N_t$ new design locations $D=\{\bm{d}_1, \dots ,\bm{d}_{N_{t}} \}$  by adaptive experimental design scheme (NFBD, NFBD-FG or NFBD-AG) with $\epsilon_0$ (Section \ref{set:NFBD}) \;
Evaluate the  corresponding   values $\{y_1, \dots ,y_{N_{t}} \}$ at $D$ by accessing the limit state function $g(\x)$\;
Use the new data-set $\mathcal{D} = \mathcal{D} \cup \{(\bm{d}_i, y_i)\}_{i=1,\dots,N_t}$ to retrain  the NN surrogate $G$\;
Estimate the failure probability $\hat{p}_f^t$ by Equation \eqref{eq:MC} with the NN surrogate $G(\x)$\;
\If  {$|\frac{\hat{P}_f^{t} - \hat{P}_f^{t-1}}{\hat{P}_f^{t-1}}| >tolerance $} 
{$\hat{P}_f = \hat{P_f^t}$ ; Break the loop;}
}

\end{algorithm}
\DecMargin{1em}
Some remarks on the implementation of algorithm \ref{alg} are listed in order:
\begin{itemize}
\item[1] In normalizing flows the choice of $p(z_0)$  can be any common probability density like uniform distribution, Gaussian distribution, etc., and in our numerical examples, we set $p(\z_0)=\mathcal{N}(\z_0;\bm{0},I)$ as the Standard Gaussian distribution without loss of generality. 
\item[2] The threshold $\epsilon_0$ is required in both NFBD-FG and NFBD-AG strategies, but unnecessary in NFBD strategy.
\item[3] The stopping criterion, $|\frac{\hat{P}_f^{t} - \hat{P}_f^{t-1}}{\hat{P}_f^{t-1}}| < tolerance $, refers that the loop can be stopped when the relative error of the failure probability obtained by two adjacent iterations is  smaller than a preset tolerance value, $torelance$.
In the numerical examples, we set $torelance = 10\%$.
\end{itemize}

\textbf{The performance of the DNF algorithm with NFBD, NFBD-FG and NFBD-AG design criteria}

As aforementioned in Section \ref{set:NFBD}, NFBD criterion is a greedy strategy without any generalization assumption and thus  DNF algorithm with NFBD criterion would lead to an accurate estimation of failure probability with the data increase. 
But This criterion would be inefficient because the proposed design points may be close together and/or be allocated in the low probability domain. 
The assumption of generalization error in both NFBD-FG and NFBD-AG criteria could avoid the situation of points close together, but would inevitably introduce a certain irreducible generalization error.
Compared with NFBD-FG, the generalization error assumed by NFBD-AG criterion would be small in high probability domain and big in low probability domain caused by its adaptive threshold function \eqref{eq:AG}.
In this setting, a relative smaller generalization assumption is given in high probability domain and bigger one in low probability domain, compared with $\epsilon_0$.

\section{Numerical examples}
\label{set:numerical_examples}

In this section we first consider the failure probability estimation problem in two $2-$d mathematical examples so that we can validate the approximate limit states with the exact ones.
 In Section \ref{set:pde}, we apply our method to a computational intensive PDE simulation problem: the
Darcy Flow equation, which is a changeling benchmark problem.

\subsection{Four branch system}

\begin{figure}
  \centering
	\includegraphics[width=.65\linewidth]{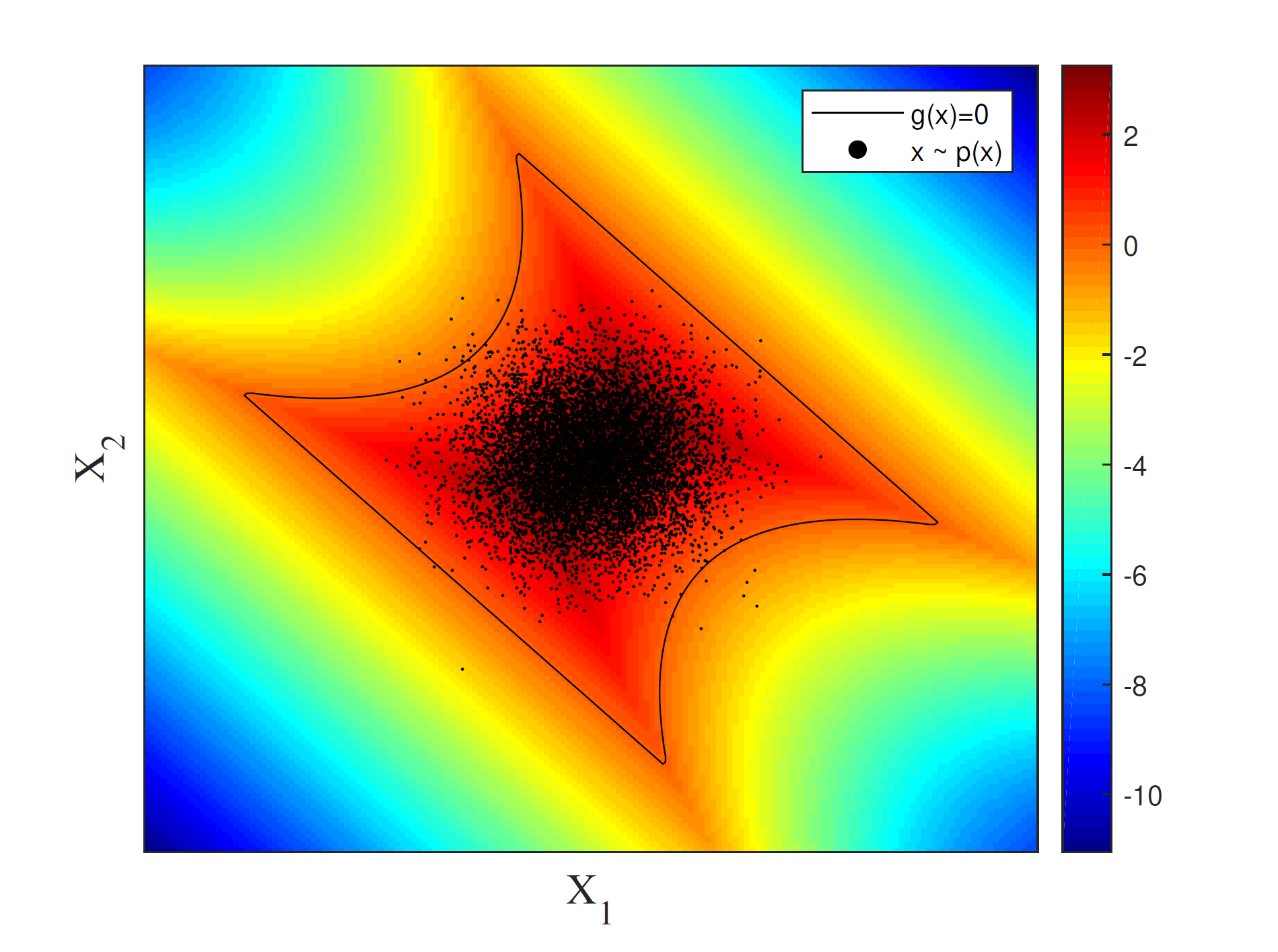}
	\caption{For four branch system example, the limit state function and the samples drawn from the prior distribution $p(\bm{x})$. The solid line indicates the failure boundary.} 
	\label{f:eg1_1}
\end{figure}
Our first example is the so-called four branch system, which is a popular benchmark test case in reliability analysis.
In this example the limit state function reads
\begin{equation}
g\left(x_1, x_2\right)=\min \left\{\begin{array}{l}
3+0.1\left(x_1-x_2\right)^2-\left(x_1+x_2\right) / \sqrt{2} \\
3+0.1\left(x_1-x_2\right)^2+\left(x_1+x_2\right) / \sqrt{2} \\
\left(x_1-x_2\right)+7 / \sqrt{2} \\
\left(x_2-x_1\right)-7 / \sqrt{2}
\end{array}\right\}
\end{equation}
which is shown in Figure \ref{f:eg1_1}.

The input random variable $x_1$ and $x_2$ are assumed to be independent and follow standard normal distribution, i.e., $p(x_i)\sim \mathcal{N}(0,1), i=1,2$.
We first compute the failure probability with a standard MC estimation of $10^5$ samples, resulting an estimate of $2.05\times 10^{-3}$.

\begin{figure}[H]
  \centering
	\includegraphics[width=.32\linewidth]{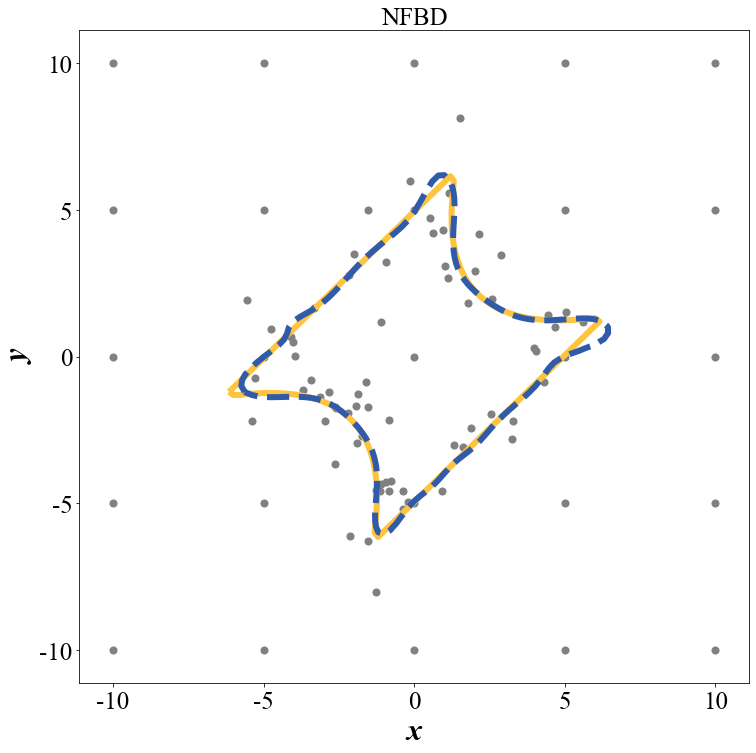}
	\includegraphics[width=.32\linewidth]{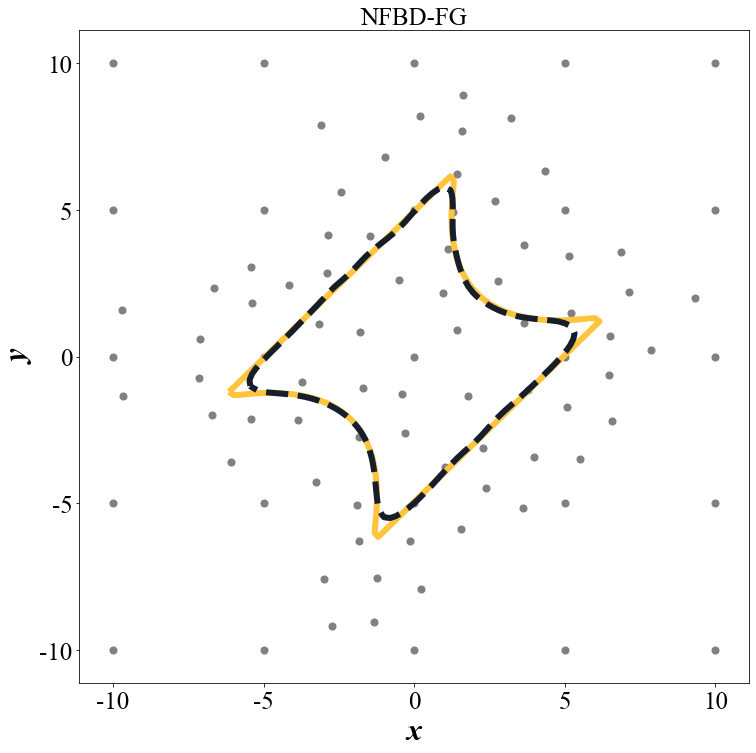}
	\includegraphics[width=.32\linewidth]{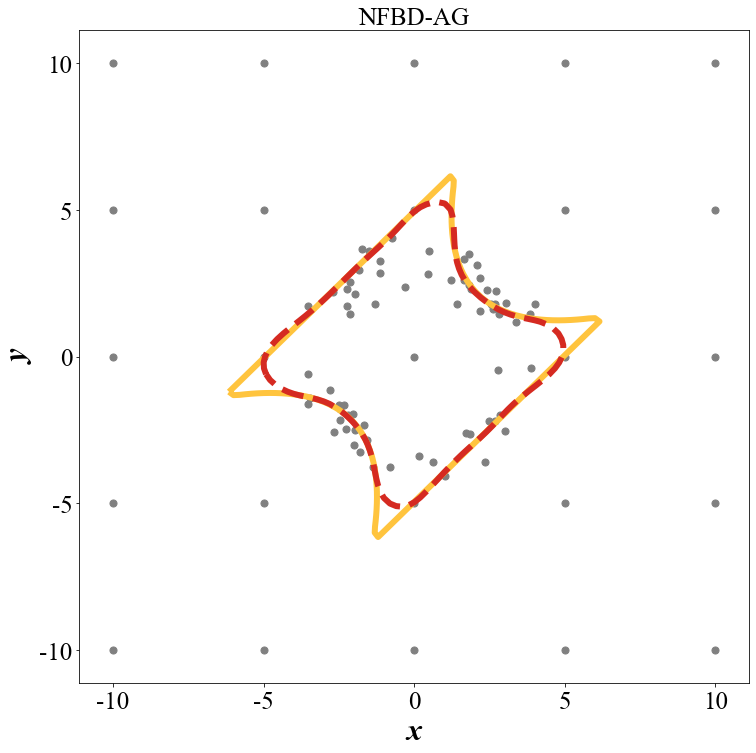}
	\caption{The gray points represent the design locations determined by NFBD (left), NFBD-FG (middle) and NFBD-AG (right) strategies, respectively. The solid line is the true limit state and the dashed line represents its approximation learned by different strategies.} 
	\label{f:eg1_3}
\end{figure}

\begin{figure}[H]
  \centering
	\includegraphics[width=.45\linewidth]{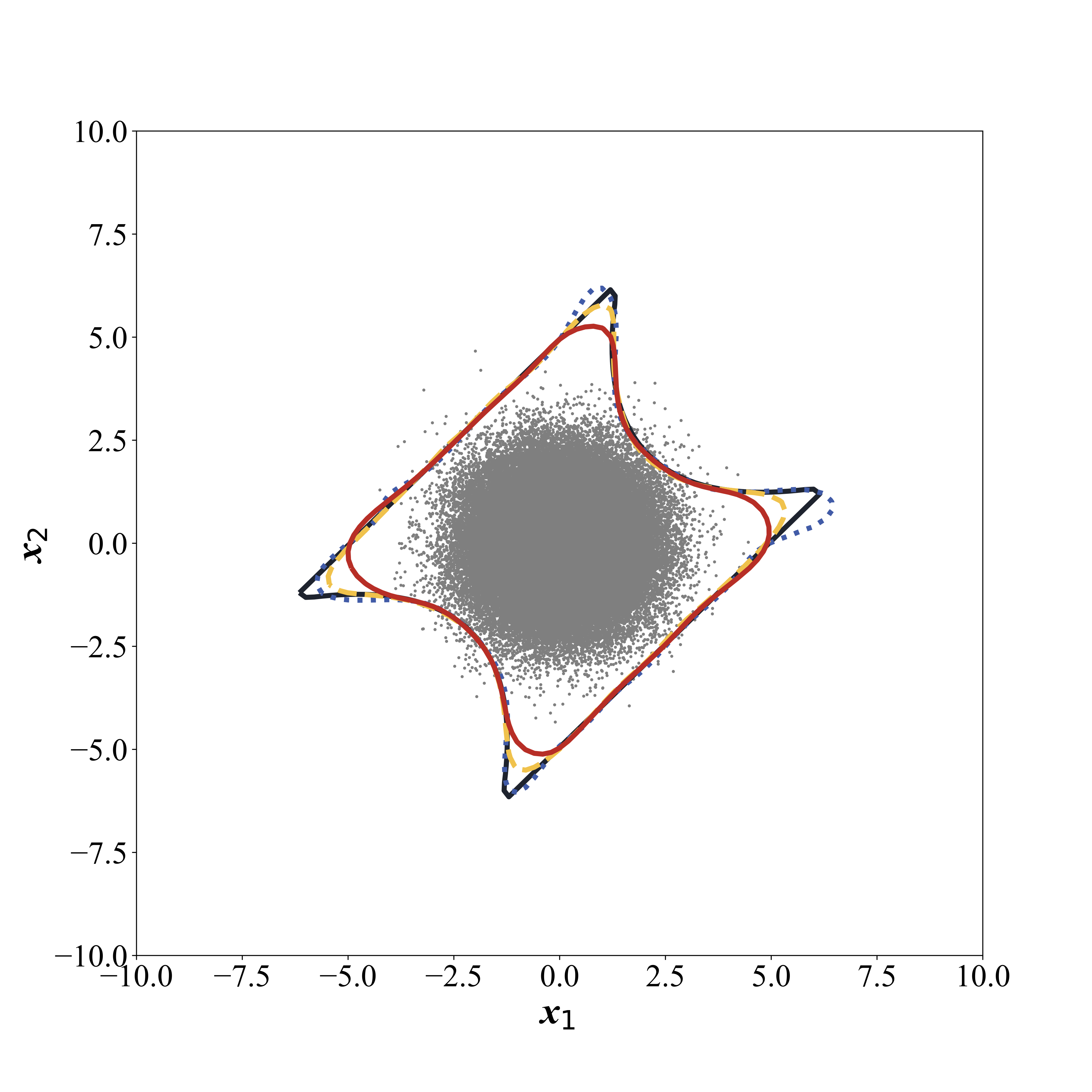}
	\includegraphics[width=.45\linewidth]{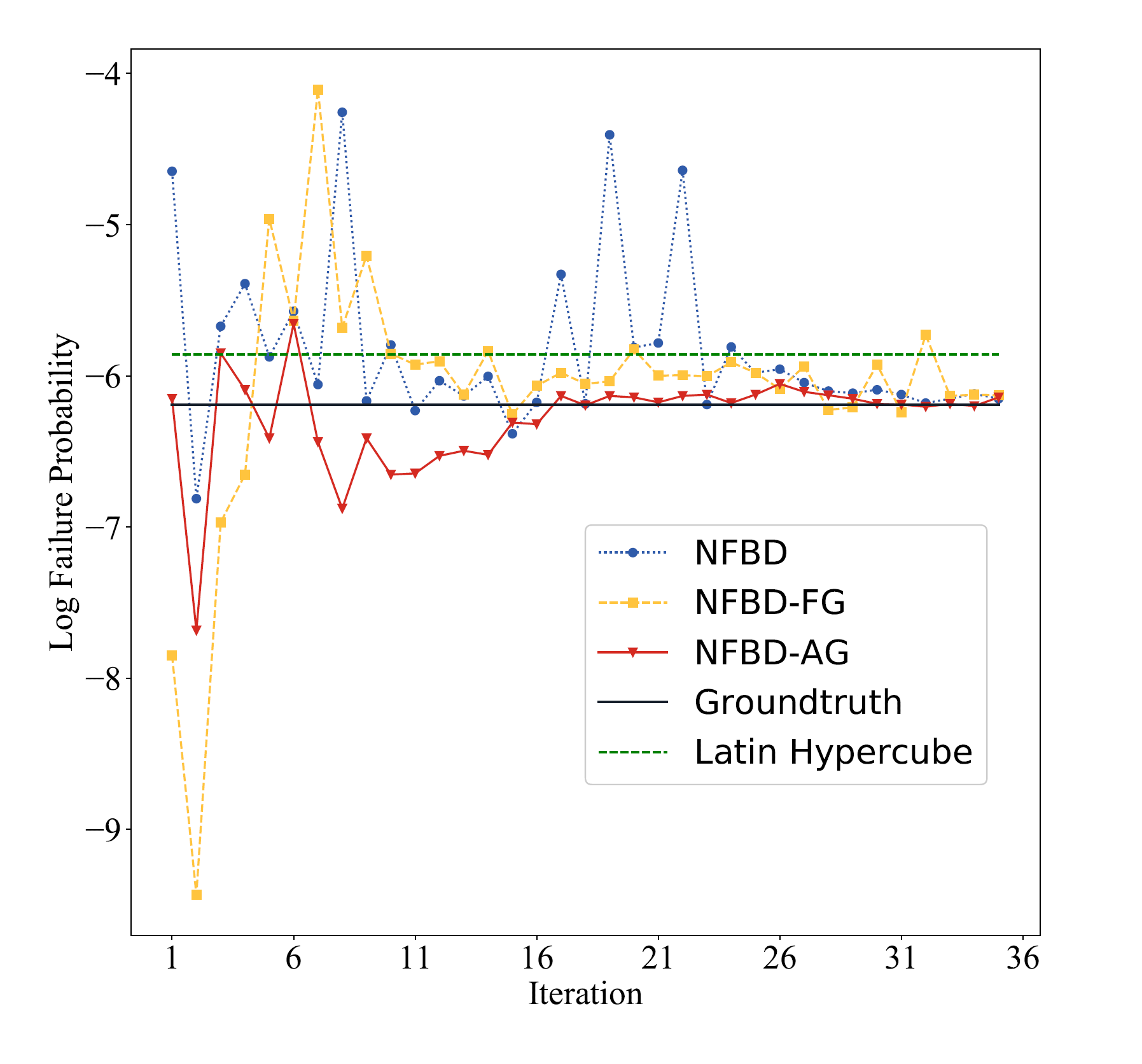}
	\caption{The true limit state (solid line) and that computed
with the NN surrogate (dashed line) are also shown in left figure. The  estimation of logarithmic failure probability vs iterations are shown in right figure.} 
	\label{f:eg1_4}
\end{figure}

In the proposed adaptive experimental design methods, we first choose $25$ points equally spaced  
in $\mathcal{X} = [-10,10]\times [-10,10]$ as the initial design points and then $2$ design points determined in each iteration.
In order to fully demonstrate the characteristics of different design criteria, we let the DNF algorithm terminates until the number of simulations reaches its max $N_{max}=95$, resulting in totally $35$ iterations. 
\begin{figure}[H]
  \centering
	\includegraphics[width=.95\linewidth]{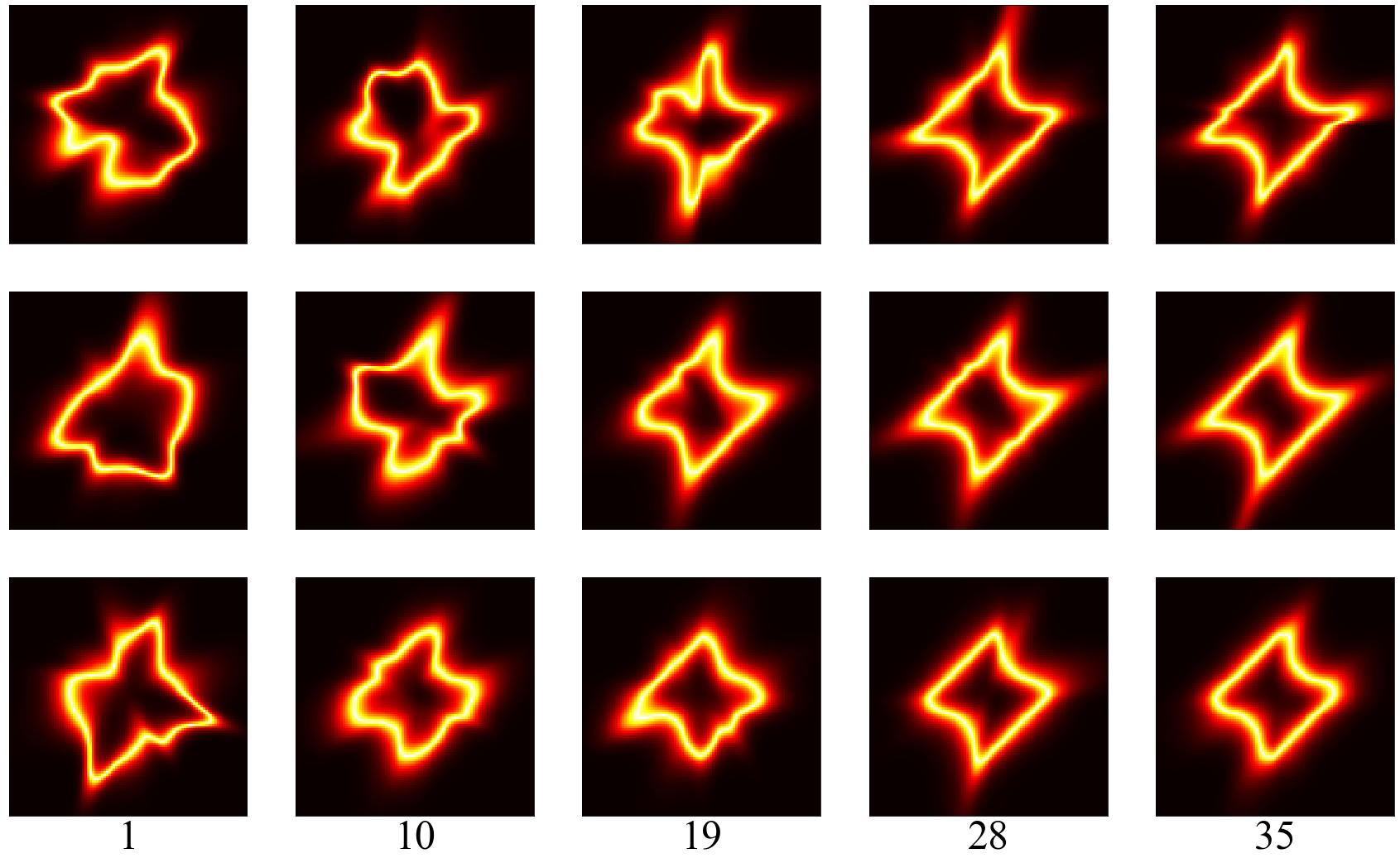}
	\caption{For four branch system example, the approximation of limit state learned by NFBD (top row), NFBD-FG (middle row) and NFBD-AG (bottom row) criteria, with respect to the first, 10th, 19th, 28th and last iteration.} 
	\label{f:eg1_2}
\end{figure}
Figure \ref{f:eg1_2} shows the approximations of limit state learned by DNF with NFBD, NFBD-FG and NFBD-AG strategies.
We can find that approximations computed by all the three criteria are gradually approaching the real one with the increase of iteration.
It verifies that the failure boundary can be obtained by our normalizing-flows-based design framework. 
We plot the design points  and their corresponding approximations of limit state computed by these three design criteria in Figure \ref{f:eg1_3}.
We can see that both NFBD and NFBD-AG methods allocate more points near the boundary of the failure domain than NFBD-FG and it is reasonable because of the fixed generalization error assumption which makes the design points more sparse in $\mathcal{X}$.
We now compare the results of the three design strategies in Figure \ref{f:eg1_4}.
The left subfigure shows that all the three methods provide an acceptable fitting accuracy in the high probability area of failure boundary.
We plot the estimation of logarithmic failure probability  as a function of the number of iterations in Figure \ref{f:eg1_4} (right).
In the figure we can see that the curve of our basis NFBD method oscillates greatly before $23$th iteration and it asymptotically stably converges to the groundtruth.
The curve of NFBD-FG method quickly stabilizes around the groundtruth, but the estimation accuracy does not increase significantly as the number of iterations increases.
It is because a fixed generalization assumption could lead to a quick exploration for the failure boundary and a upper limit of  the accuracy of this estimation.
Compare with the other two methods, the curve of NFBD-AG performs well both on the speed of stabilization and the estimation accuracy, which benefits from its adaptive generalization error.
The estimates of failure probability computed by NFBD, NFBD-FG and NFBD-AG in the last iteration are $2.13\times 10^{-3}$, $2.18\times 10^{-3}$ and $2.15 \times 10^{-3}$, respectively, which are more accurate than the one $2.86\times 10^{-3}$ computed by Latin hypercube sampling method with the same number of evaluations.
We can find that all these criteria result in a less than ten percent relative error.

\subsection{Iso-probability lines}
Our second example is the iso-probability lines and its specific expression of the system is
\begin{equation}
    g(x_1,x_2) = b - x_2 - k(x_1-e)^2,
\end{equation}
where $b=5$, $k=0.5$ and $e=0.1$. 
The prior of each parameter in $\x=[x_1, x_2]$ is also standard Normal distribution.
$10^5$ samples are used for MC estimation of the failure probability, resulting an estimate of $3.01\times 10^{-3}$.
\begin{figure}[H]
  \centering
	\includegraphics[width=.65\linewidth]{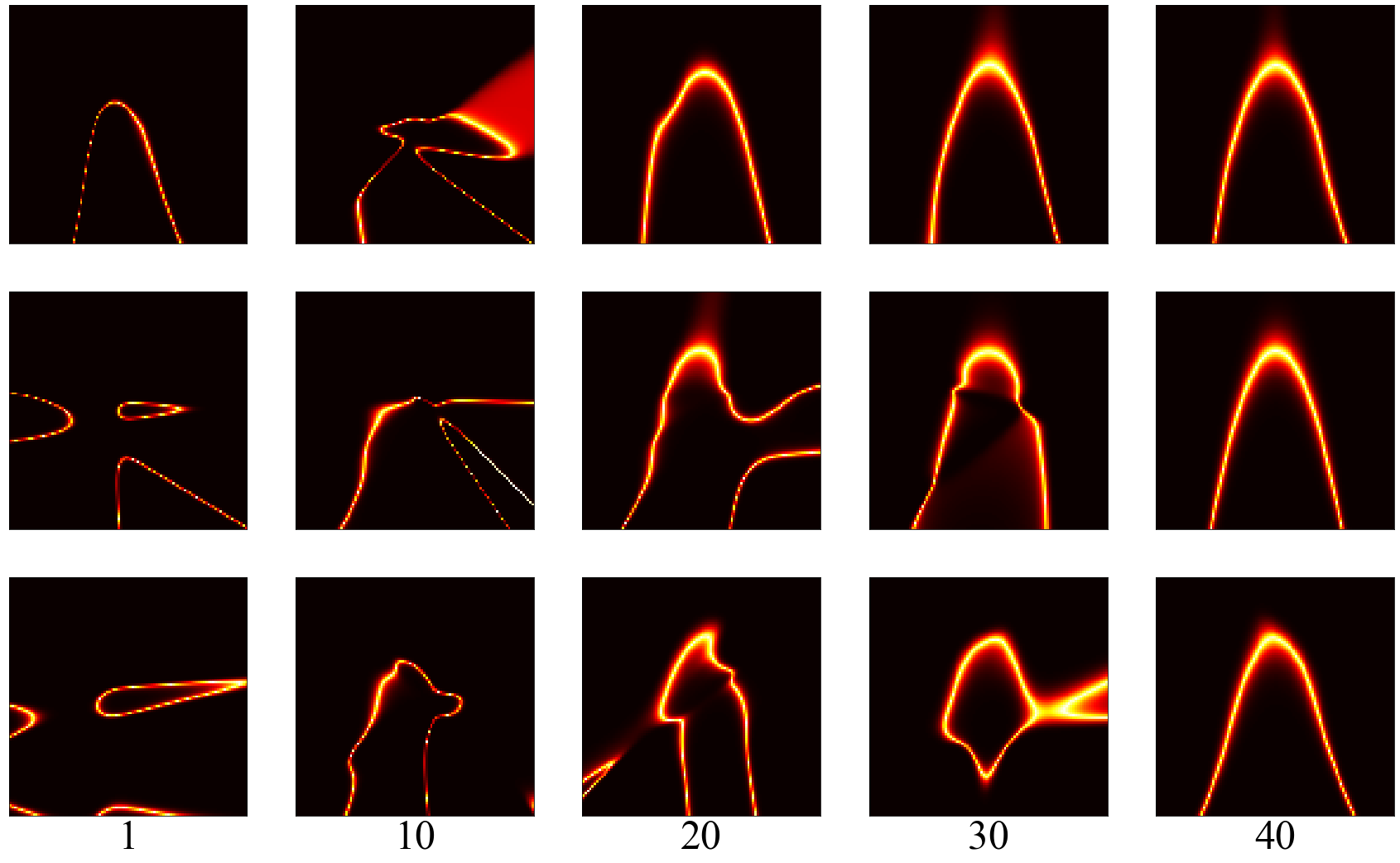}
	\caption{For iso-probability lines example,  the approximations of limit state learned by NFBD (top row), NFBD-FG (middle row) and NFBD-AG (bottom row) criteria, with respect to the first, 10th, 20th, 30th and last iterations, respectively.} 
	\label{f:eg2_1}
\end{figure}

In this example, $5$ points equally spaced in $\mathcal{X} = [-10,10]\times[-10,10]$ are chosen as the initial design points and $40$ design points are determined by the DNF algorithm with one point determined in each iteration.

\begin{figure}[H]
  \centering
	\includegraphics[width=.32\linewidth]{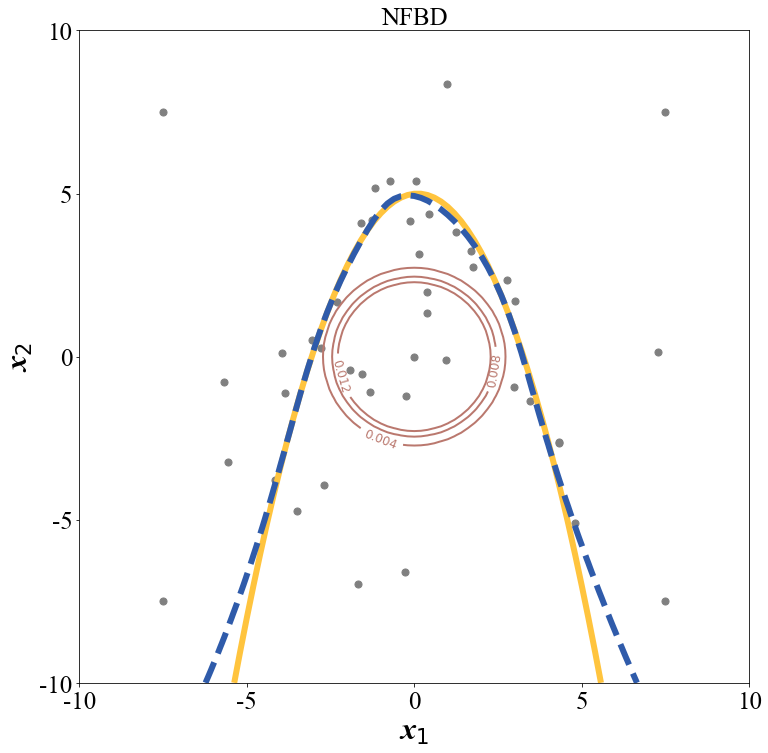}
	\includegraphics[width=.32\linewidth]{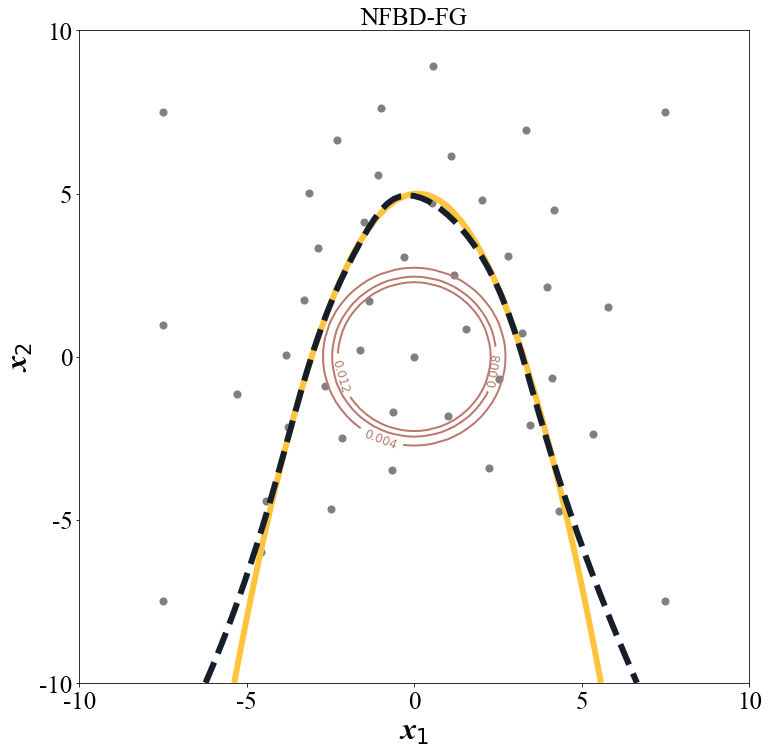}
	\includegraphics[width=.32\linewidth]{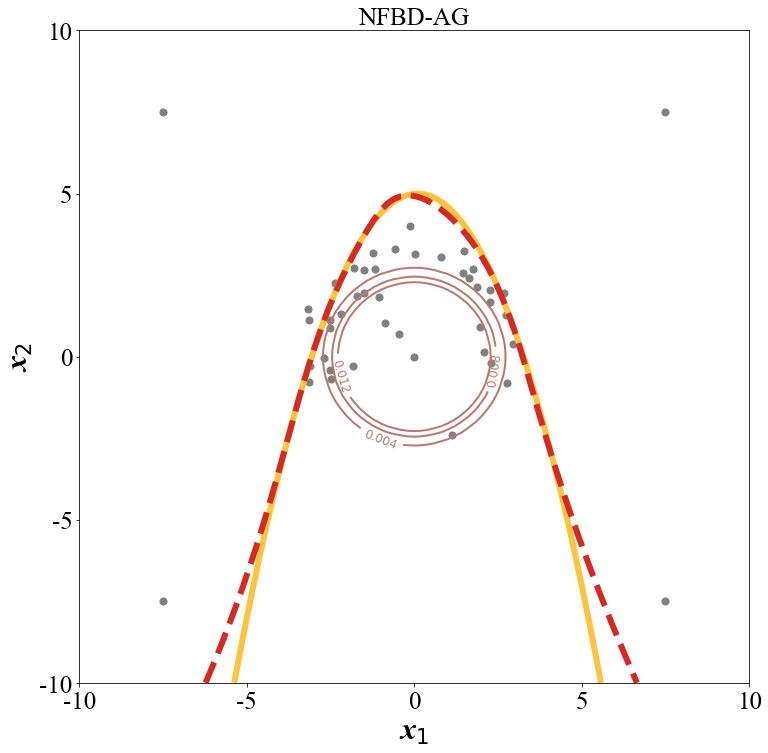}
	\caption{For iso-probability lines example, the design locations (gray dots) and the approximation of limit state (dashed line) learned by NFBD, NFBD-FG and NFBD-AG, respectively. The contour lines represent the Gaussian distribution of $p(\x)$. } 
	\label{f:eg2_2}
\end{figure}

We plot the approximate posteriors obtained in the first, 10th, 20th, 30th and last iterations in Figure \ref{f:eg2_1}, in which we can visualize how the quality of the approximation increases as the iterations proceed.
We can find that all the approximate posterior distributions of limit state in the first twenty iterations are far from the real one and they are gradually approaching the real one ( as shown in Figure \ref{f:eg2_2}) in the end with the increase of iteration.
Figure \ref{f:eg2_2} shows all the design points determined by DNF with different criteria and we can find that the design samples computed by NFBD-AG allocates more points near the failure boundaries where larger PDF value of $p(\x)$ has.
The distribution of these design points are different from the one computed by NFBD which allocates the more design points around the boundary of failure, and the one computed by NFBD-FG which sparsely allocates the design points around the failure boundary.
In theory, NFBD-AG allocates the points with considering the value of $p(\tilde{\z}|\mathcal{D})$,i.e., the pdf value of the approximate posterior of limit state and  $p(\x)$.

\begin{figure}[H]
  \centering
	\includegraphics[width=.65\linewidth]{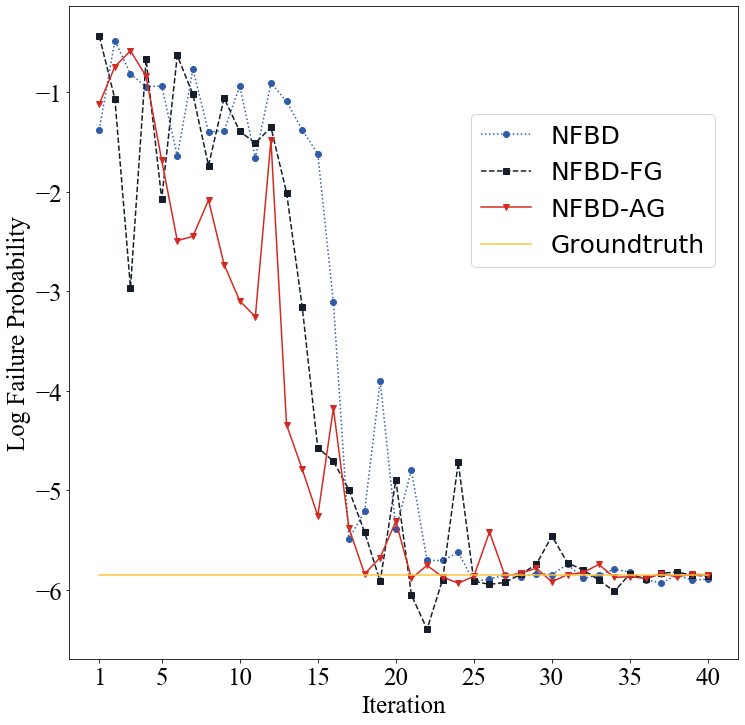}
	\caption{For iso-probability lines example, the estimation of logarithmic failure probability  vs iterations.} 
	\label{f:eg2_3}
\end{figure}

Figure \ref{f:eg2_3} shows the estimation of  logarithmic failure probability as a function of the number of iterations.
We can see that the plot of NFBD-AG has the fastest convergence speed and a more stable performance after 27th iteration.
The plot of NFBD has a slower convergence speed and the plot of NFBD-FG has a more unstable representation after convergence, compared with the plot of NFBD-AG.
These results verified the statement that NFBD has a slower convergence speed and NFBD-FG has a faster convergence speed with a large generalization error as mentioned in Section \ref{set:alg}.
In the last iteration, the estimates of failure probability computed by NFBD, NFBD-FG and NFBD-AG converge to $2.76\times 10^{-3}$, $2.84\times 10^{-3}$ and $2.83\times 10^{-3}$, respectively, which are more accurate than the one $2.69\times 10^{-3}$ computed by Latin hypercube sampling method with the same number of evaluations.
The relative error computed by all these estimations are less than ten percent.

\subsection{PDE problem cases}
\label{set:pde}

Particularly, We pay a attention on a specific scenario where  the failure probability estimation  of the parametric two-dimensional Darcy Flow equation. The input of this PDE is a function which called drift function and its output is the solution of PDE. 
The $2$-d Darcy Flow equation is defined on the unit box which is the second order, linear, elliptic PDE

\begin{equation}
    \begin{array}{cc}
         -\nabla \cdot (a(\bm{\xi};\x)\nabla u(\bm{\xi}))&=f(\bm{\xi})\quad  \bm{\xi}\in(0,1)^2\\
         u(\bm{\xi})&= 0 \quad  \bm{\xi}\in \partial (0,1)^2
    \end{array}
\end{equation}
with a Dirichlet boundary where $a\in L^{\infty}((0,1)^2\times \mathbb{R}^d;\mathbb{R}_{+})$ is the diffusion coefficient parametric by $\x \in \mathbb{R}^d$ and $f \in L^2((0,1)^2;\mathbb{R})$ is the forcing function which is given $f(\bm{\xi}) = 1$. 

\begin{figure}[H]
  \centering
	\includegraphics[width=.45\linewidth]{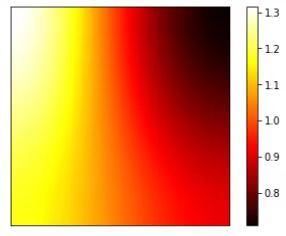}
	\includegraphics[width=.45\linewidth]{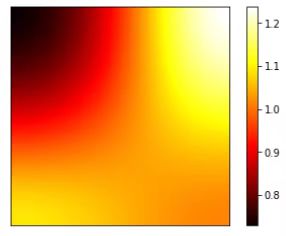}
	\caption{For PDE problem example, two samples of $a(\bm{\xi})$ drawn from $\mu$ .} 
	\label{f:exam3_a}
\end{figure}

We are interest in the situation when $a(\bm{\xi};\x)$ belong to a exponential random field, i.e. is random function, the solution $u(\bm{\xi})$ would be random and determined by the random parameter $\x$.
The coefficients $a(\bm{\xi};\x)$ are generated according to $\ln a\sim \mu$ where $\mu=\mathcal{N}(\mu_{\ln a}, \sigma_{\ln a})$, where $\mu_{\ln a}=0$ and $\sigma_{\ln a}=(-\triangle + 9I)^{-2})$ with zero Neumann boundary conditions on the Laplacian. 
To represent the random field $\ln a(\bm{\xi};\x)$, we apply its Karhunen-Loeve expansion which takes the following form 
\begin{equation*}
\ln a(\xi) = \mu_{\ln a} + \sigma_{\ln a}\sum^d_{i=1}\sqrt{\theta_i}\mathcal{X}_i(\xi)x_i
\end{equation*}
where $\{\theta_i,\mathcal{X}_i\}$ are the eigenpairs of $\rho_{\ln a}$, which are known analytically for the applied exponential correlation model, and $x_i, i=1,\dots,d$ are independent standard normal random variables.
The expansion is truncated after $4$ terms, i.e., $d=4$ and Figure \ref{f:exam3_a} shows two samples of $a(\bm{\xi})$.
We employ this discrete representation of the log-diffusivity and used to define the failure event.
Then the limit state function is defined as:
\begin{equation*}
D_{\epsilon}(u(\x)) = \max(u(\x)) - \epsilon
\end{equation*} 
where the threshold $\epsilon = 0.082$.

\begin{figure}[H]
  \centering
	\includegraphics[width=.65\linewidth]{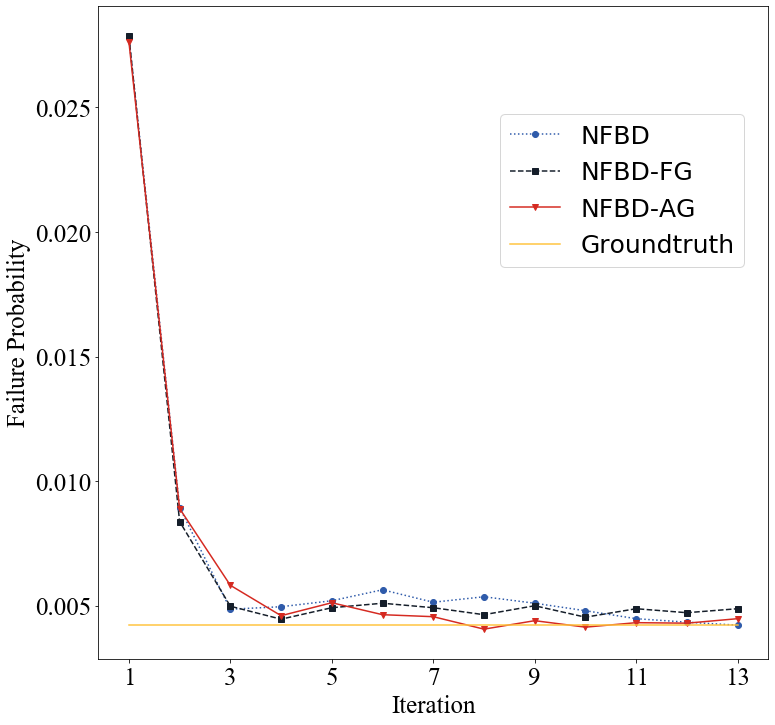}
	\caption{For PDE problem example, the estimation of failure probability  vs iterations.} 
	\label{f:exam3_1}
\end{figure}

In this example, we draw $5$ initial design points by Latin hypercube sampling (LHS) method and $65$ design points are determined by the DNF algorithm with $5$ design points in each iteration.
The estimations of failure probability computed by different criteria,  with respect to different iterations, are shown in Figure \ref{f:exam3_1}.
We can find all the estimations of the proposed three criteria successfully converge to the true value $4.24\times 10^{-3}$.
Same as the previous two examples, the plot of NFBD slowly and accurately converge to the real value and the the plot of NFBD-FG converge fast but not accurately.
The estimation error of NFBD-FG are not improved with the increase of iteration.
The curve of NFBD-AG quickly and smoothly converges to the real one.
In the last iteration, the estimates of failure probability computed by NFBD, NFBD-FG and NFBD-AG converge to $4.22\times 10^{-3}$, $4.88\times 10^{-3}$ and $4.88\times 10^{-3}$, respectively, which are more accurate than the one $5.42\times 10^{-3}$ computed by Latin hypercube sampling method with the same number of evaluations.
The relative error computed by all these estimations are less than ten percent.

\section{Conclusion}\label{conclusion}
\label{set:conclusion}

In conclusion, we have presented an adaptive design via normalizing flows (DNF) scheme for failure boundary detection and neural network model as the surrogate of computational intensive simulation for failure probability estimations.
In particular, the method recasts the failure detection as inferring a contour $g(\x)=0$ with Bayesian methods,
and then adaptively determines the design locations  for the inference problem.
Three normalizing-flows-based design criteria, NFBD, NFBD-FG and NFBD-AG, are also presented. 
With numerical examples, we demonstrate that the proposed method can efficiently determine design points for failure detection and failure probability estimation.

\section*{Acknowledgement}
  Hongqiao Wang acknowledges the support of NSFC 12101615 and  the Natural Science Foundation of Hunan Province, China, under Grant 2022JJ40567.
  This work was carried out in part using computing resources at the High Performance Computing Center of Central South University.
\appendix
\section*{Appendices}
\section{Fourier Neural Operator}
\label{app:FNO}
FNO methodology learns a mapping between two infinite dimentional spaces from a finite collection of observed input-output pairs.
Let $D\subset\mathbb{R}^d$ be a bounded, open set and $\mathcal{A}=\mathcal{A}(D;\mathbb{R}^{d_a})$ and $\mathcal{U}=\mathcal{U}(D;\mathbb{R}^{d_{u}})$ be separable Banach spaces of function taking values in $\mathbb{R}^{d_a}$ and $\mathbb{R}^{d_u}$ respectively.
Furthermore let $g^\dagger:\mathcal{A}\to \mathcal{U}$ be a (typically) non-linear map which arises as the solution operators of parametric PDEs, such as the 2-d Darcy Flow equation,
\begin{equation}
    \begin{array}{cc}
         -\nabla \cdot (a(\bm{\xi})\nabla u(\bm{\xi}))&=f(\bm{\xi})\quad  \bm{\xi}\in(0,1)^2\\
         u(\bm{\xi})&= 0 \quad  \bm{\xi}\in \partial (0,1)^2.
    \end{array}
\end{equation}
Suppose we have the observations $\{a_j,u_j \}^N_{j=1}$ where $a_j \sim \mu$ is an i.i.d. sequence from the probability measure $\mu$ supported on $\mathcal{A}$ and $u_j = g^\dagger(a_j)$ is possibly corrupted with noise.
We aim to build an approximation of $g^\dagger$ by constructing a parametric map
\begin{equation}
    G:\mathcal{A}\times \Theta \to \mathcal{U}\quad \text{or equivalently,}\quad G_\theta:\mathcal{A} \to \mathcal{U}, \theta\in\Theta
\end{equation}
for some finite-dimensional parameter space $\Theta$ by choosing $\theta^* \in \Theta$ so that $G(\cdot,\theta^*)=G_{\theta^*}\approx g^\dagger$.
This is a natural framework for learning in infinite-dimensions as one could define a cost functional $C:\mathcal{U}\times \mathcal{U}\to \mathbb{R}$ and seek a minimizer of the problem 
\begin{equation}
    \min_{\theta\in \Theta}\mathbb{E}_{a\sim \mu}[C(G(a,\theta),G^\dagger(a))]
\end{equation}
which directly parallels the classical finite-dimensional setting.
We will approach this problem in the test-train setting by using a data-driven empirical approximation to the cost used to determine $\theta$ and to test the accuracy of the approximation. Details about FNO please refer \cite{li2020fourier}.


\section{Normalizing Flows}
\label{app:NF}
Here we give a brief introduction of Normalizing Flows and more details please refer \cite{rezende2015variational}.
The basic rule for transformation of densities considers an invertible, smooth mapping $f:\mathbb{R}^d\to \mathbb{R}^d$ with inverse $f^{-1}=h$, i.e. the composition $h \circ f(\z_0) = \z_0$. 
If we use this mapping to transform a random variable $\z_0$ with distribution $q(\z_0)$, the resulting random variable $\z=f(\z_0)$ has a distribution:
\begin{equation}
\label{eq:NF_flow}
q(\z)=q(\z_0)|\det \frac{\partial f^{-1}}{\partial \z}| = q(\z_0)|\det \frac{\partial f}{\partial \z_0}|^{-1},
\end{equation}
where the last equality can be seen by applying the chain rule (inverse function theorem) and is a property of Jacobians of invertible functions.
We can construct arbitrarily
complex densities by composing several simple maps and
successively applying Equation \eqref{eq:NF_flow}.
The density $q_K(\z_K)$ obtained by successively transforming a random variable $\z_0$ with distribution $q_0$ through a chain of $K$ transformations $f_k$ is :
\begin{align}
\label{eq:NF_composition}
\z_K &= f_K\circ\dots\circ f_2\circ f_1(\z_0)\\
\ln q_K(\z_K) &= \ln q_0(\z_0) - \sum^K_{k=1}\ln \det|\frac{\partial f_k}{\partial \z_k}|,
\end{align}
where equation \eqref{eq:NF_composition} will be used as a shorthand for the composition $f_K(f_{K-1}(\dots f_1(\z)))$.
The path traversed by the random variables $\z_k=f_k(\z_{k-1})$ with initial distribution $q_0(\z_0)$ is called the flow and the path formed by the successive distributions $q_k$ is a normalizing flow.

Consider a general probabilistic model with observation $\y$, latent variables $\z$ over which we must integrate, and model parameters $\btheta$.
We introduce an approximate posterior distribution for the latent variable $q_\phi(\z|\y)$ and follow the variational principle \cite{jordan1999introduction} to obtain a bound on the marginal likelihood:
\begin{align}
\label{eq:Jensen}
\ln p_\theta(\y) &= \ln \int p_\theta (\y|\z) p(\z)d\z\\
&=\ln\int \frac{q_\phi(\z|\y)}{q_\phi(\z|\y)}p_\theta(\y|\z)p(\z)d\z\\
&\ge \mathbb{D}_{KL}[q_\phi(\z|\y)||p(\z)] + \mathbb{E}_q[\ln p_\theta(\y|\z)]=-\mathcal{F}(\y),
\end{align}
where we used Jensen's inequality to obtain the final equation, $\mathbb{D}_{KL}[\cdot||\cdot]$ is Kullback-Leibler Divergence,  $p_\theta(\y|\z)$ is a likelihood function and $p(\z)$ is a prior over the latent variables.

If we parameterize the approximate posterior distribution with a flow of length $K$, $q_{\phi}(\z|\y) := q_K(z_K)$, the free energy can be  \eqref{eq:Jensen} can be written as an expectation over the initial distribution $q_0(\z_0)$:
\begin{equation}
\label{eq:freeEnergy}
\begin{aligned}
\mathcal{F}(\y) &=\mathbb{E}_{q_\phi(\z \mid \y)}\left[\ln q_\phi(\mathbf{z} \mid \y)-\ln p(\y, \mathbf{z})\right] \\
&=\mathbb{E}_{q_0\left(\z_0\right)}\left[\ln q_K\left(\mathbf{z}_K\right)-\ln p\left(\y, \mathbf{z}_K\right)\right] \\
&=\mathbb{E}_{q_0\left(\z_0\right)}\left[\ln q_0\left(\mathbf{z}_0\right)\right]-\mathbb{E}_{q_0\left(z_0\right)}\left[\ln p\left(\y, \mathbf{z}_K\right)\right] \\
&-\mathbb{E}_{q_0\left(\z_0\right)}\left[\sum_{k=1}^K \ln \det|\frac{\partial f_{k,\phi}}{\partial \z_k}|\right]
\end{aligned}.
\end{equation}
The neural network based transform function $f_{k,\phi}$ can be obtained by maximizing the free energy in Equation \eqref{eq:freeEnergy}.

\bibliography{FB_NF}

\begin{thebibliography}{10}

\bibitem{abd2021advanced}
Mohamed Abd~Elaziz, Abdelghani Dahou, Laith Abualigah, Liyang Yu, Mohammad
  Alshinwan, Ahmad~M Khasawneh, and Songfeng Lu.
\newblock Advanced metaheuristic optimization techniques in applications of
  deep neural networks: a review.
\newblock {\em Neural Computing and Applications}, 33(21):14079--14099, 2021.

\bibitem{au2001estimation}
Siu-Kui Au and James~L Beck.
\newblock Estimation of small failure probabilities in high dimensions by
  subset simulation.
\newblock {\em Probabilistic engineering mechanics}, 16(4):263--277, 2001.

\bibitem{bect2012sequential}
Julien Bect, David Ginsbourger, Ling Li, Victor Picheny, and Emmanuel Vazquez.
\newblock Sequential design of computer experiments for the estimation of a
  probability of failure.
\newblock {\em Statistics and Computing}, 22(3):773--793, 2012.

\bibitem{bichon2008efficient}
Barron~J Bichon, Michael~S Eldred, Laura~Painton Swiler, Sandaran Mahadevan,
  and John~M McFarland.
\newblock Efficient global reliability analysis for nonlinear implicit
  performance functions.
\newblock {\em AIAA journal}, 46(10):2459--2468, 2008.

\bibitem{blum2010non}
Michael~GB Blum and Olivier Fran{\c{c}}ois.
\newblock Non-linear regression models for {A}pproximate {B}ayesian
  {C}omputation.
\newblock {\em Statistics and computing}, 20(1):63--73, 2010.

\bibitem{chen2019adaptive}
Yanzhi Chen and Michael~U Gutmann.
\newblock Adaptive {G}aussian copula {ABC}.
\newblock In {\em The 22nd International Conference on Artificial Intelligence
  and Statistics}, pages 1584--1592. PMLR, 2019.

\bibitem{chevalier2014fast}
Cl{\'e}ment Chevalier, Julien Bect, David Ginsbourger, Emmanuel Vazquez, Victor
  Picheny, and Yann Richet.
\newblock Fast parallel kriging-based stepwise uncertainty reduction with
  application to the identification of an excursion set.
\newblock {\em Technometrics}, 56(4):455--465, 2014.

\bibitem{dubourg2013metamodel}
Vincent Dubourg, Bruno Sudret, and Franois Deheeger.
\newblock Metamodel-based importance sampling for structural reliability
  analysis.
\newblock {\em Probabilistic Engineering Mechanics}, 33:47--57, 2013.

\bibitem{echard2011ak}
Benjamin Echard, Nicolas Gayton, and Maurice Lemaire.
\newblock Ak-mcs: an active learning reliability method combining kriging and
  monte carlo simulation.
\newblock {\em Structural Safety}, 33(2):145--154, 2011.

\bibitem{engelund1993benchmark}
Svend Engelund and Ruediger Rackwitz.
\newblock A benchmark study on importance sampling techniques in structural
  reliability.
\newblock {\em Structural safety}, 12(4):255--276, 1993.

\bibitem{faravelli1989response}
Lucia Faravelli.
\newblock Response-surface approach for reliability analysis.
\newblock {\em Journal of Engineering Mechanics}, 115(12):2763--2781, 1989.

\bibitem{gayton2003cq2rs}
Nicolas Gayton, Jean~Marc Bourinet, and Maurice Lemaire.
\newblock Cq2rs: a new statistical approach to the response surface method for
  reliability analysis.
\newblock {\em Structural safety}, 25(1):99--121, 2003.

\bibitem{jordan1999introduction}
Michael~I Jordan, Zoubin Ghahramani, Tommi~S Jaakkola, and Lawrence~K Saul.
\newblock An introduction to variational methods for graphical models.
\newblock {\em Machine learning}, 37(2):183--233, 1999.

\bibitem{jumper2021highly}
John Jumper, Richard Evans, Alexander Pritzel, Tim Green, Michael Figurnov,
  Olaf Ronneberger, Kathryn Tunyasuvunakool, Russ Bates, Augustin dek, Anna
  Potapenko, et~al.
\newblock Highly accurate protein structure prediction with alphafold.
\newblock {\em Nature}, 596(7873):583--589, 2021.

\bibitem{kabir2018neural}
HM~Dipu Kabir, Abbas Khosravi, Mohammad~Anwar Hosen, and Saeid Nahavandi.
\newblock Neural network-based uncertainty quantification: A survey of
  methodologies and applications.
\newblock {\em IEEE access}, 6:36218--36234, 2018.

\bibitem{li2011efficient}
Jing Li, Jinglai Li, and Dongbin Xiu.
\newblock An efficient surrogate-based method for computing rare failure
  probability.
\newblock {\em Journal of Computational Physics}, 230(24):8683--8697, 2011.

\bibitem{li2012bayesian}
Ling Li, Julien Bect, and Emmanuel Vazquez.
\newblock Bayesian subset simulation: a kriging-based subset simulation
  algorithm for the estimation of small probabilities of failure.
\newblock {\em arXiv preprint arXiv:1207.1963}, 2012.

\bibitem{li2020fourier}
Zongyi Li, Nikola Kovachki, Kamyar Azizzadenesheli, Burigede Liu, Kaushik
  Bhattacharya, Andrew Stuart, and Anima Anandkumar.
\newblock Fourier neural operator for parametric partial differential
  equations.
\newblock {\em arXiv preprint arXiv:2010.08895}, 2020.

\bibitem{liu2017survey}
Weibo Liu, Zidong Wang, Xiaohui Liu, Nianyin Zeng, Yurong Liu, and Fuad~E
  Alsaadi.
\newblock A survey of deep neural network architectures and their applications.
\newblock {\em Neurocomputing}, 234:11--26, 2017.

\bibitem{lydia2019adagrad}
Agnes Lydia and Sagayaraj Francis.
\newblock Adagrad—an optimizer for stochastic gradient descent.
\newblock {\em Int. J. Inf. Comput. Sci}, 6(5):566--568, 2019.

\bibitem{oakley2002bayesian}
Jeremy Oakley and Anthony O'Hagan.
\newblock Bayesian inference for the uncertainty distribution of computer model
  outputs.
\newblock {\em Biometrika}, 89(4):769--784, 2002.

\bibitem{papamakarios2019sequential}
George Papamakarios, David Sterratt, and Iain Murray.
\newblock Sequential neural likelihood: Fast likelihood-free inference with
  autoregressive flows.
\newblock In {\em The 22nd International Conference on Artificial Intelligence
  and Statistics}, pages 837--848. PMLR, 2019.

\bibitem{raissi2019physics}
Maziar Raissi, Paris Perdikaris, and George~E Karniadakis.
\newblock Physics-informed neural networks: A deep learning framework for
  solving forward and inverse problems involving nonlinear partial differential
  equations.
\newblock {\em Journal of Computational physics}, 378:686--707, 2019.

\bibitem{renganathan2022multifidelity}
Ashwin Renganathan, Vishwas Rao, and Ionel Navon.
\newblock Multifidelity gaussian processes for failure boundary and probability
  estimation.
\newblock In {\em AIAA SCITECH 2022 Forum}, page 0390, 2022.

\bibitem{rezende2015variational}
Danilo Rezende and Shakir Mohamed.
\newblock Variational inference with normalizing flows.
\newblock In {\em International conference on machine learning}, pages
  1530--1538. PMLR, 2015.

\bibitem{rubinstein2004applications}
Reuven~Y Rubinstein and Dirk~P Kroese.
\newblock Applications of ce to machine learning.
\newblock In {\em The Cross-Entropy Method}, pages 251--270. Springer, 2004.

\bibitem{SCHUELLER2004463}
G.I. Schuëller, H.J. Pradlwarter, and P.S. Koutsourelakis.
\newblock A critical appraisal of reliability estimation procedures for high
  dimensions.
\newblock {\em Probabilistic Engineering Mechanics}, 19(4):463--474, 2004.

\bibitem{tripathy2018deep}
Rohit~K Tripathy and Ilias Bilionis.
\newblock Deep uq: Learning deep neural network surrogate models for high
  dimensional uncertainty quantification.
\newblock {\em Journal of computational physics}, 375:565--588, 2018.

\bibitem{wang2019nearest}
Haoyu Wang, Yawen Guan, and Brain Reich.
\newblock Nearest-neighbor neural networks for geostatistics.
\newblock In {\em 2019 international conference on data mining workshops
  (ICDMW)}, pages 196--205. IEEE, 2019.

\bibitem{wang2016gaussian}
Hongqiao Wang, Guang Lin, and Jinglai Li.
\newblock Gaussian process surrogates for failure detection: A bayesian
  experimental design approach.
\newblock {\em Journal of Computational Physics}, 313:247--259, 2016.

\bibitem{wang2015cross}
Hui Wang and Xiang Zhou.
\newblock A cross-entropy scheme for mixtures.
\newblock {\em ACM Transactions on Modeling and Computer Simulation (TOMACS)},
  25(1):1--20, 2015.

\bibitem{wang2022generalization}
Mingze Wang and Chao Ma.
\newblock Generalization error bounds for deep neural networks trained by sgd.
\newblock {\em arXiv preprint arXiv:2206.03299}, 2022.

\bibitem{zou2019sufficient}
Fangyu Zou, Li~Shen, Zequn Jie, Weizhong Zhang, and Wei Liu.
\newblock A sufficient condition for convergences of adam and rmsprop.
\newblock In {\em Proceedings of the IEEE/CVF Conference on computer vision and
  pattern recognition}, pages 11127--11135, 2019.

\end{thebibliography}
\bibliographystyle{plain} 
\end{document}